# Latent Mappings: Generating Open-Ended Expressive Mappings Using Variational Autoencoders

**Tim Murray-Browne[1], Panagiotis Tigas[2]**

[1]Preverbal Studio, [2]University of Oxford

**ABSTRACT**

In many contexts, creating mappings for gestural interactions can form part of an artistic process. Creators seeking a mapping that is expressive, novel, and affords them a sense of authorship may not know how to program it up in a signal processing patch. Tools like Wekinator [1] and MIMIC [2] allow creators to use supervised machine learning to learn mappings from example input/output pairings. However, a creator may know a good mapping when they encounter it yet start with little sense of what the inputs or outputs should be. We call this an *open-ended mapping process*. Addressing this need, we introduce the *latent mapping*, which leverages the latent space of an unsupervised machine learning algorithm such as a Variational Autoencoder trained on a corpus of unlabelled gestural data from the creator. We illustrate it with *Sonified Body*, a system mapping full-body movement to sound which we explore in a residency with three dancers.

## Introduction

The mapping defines the relationship between sensor and synthesizer in a digital musical interface. We introduce the *latent mapping*, an approach to generating arbitrary yet expressive mappings using unsupervised machine learning (ML).

Our target use cases are mappings generated as part of a creative process, such as those created for specific performative work or interactive sound installation [3], although our approach may be of value to those designing Digital Musical Instruments intended for general use. We use the term *creator* to describe the person creating the mapping, who may be a musician, composer, artist, or otherwise.

We developed this approach in response to the needs of our own artistic work transforming full-body movement to sound and we describe it in this context. However, it responds more to the needs of a certain type of creative process rather than a specific performative domain. We offer it as a general approach to creating mappings for gestural interaction.

## The Mapping Process

Mappings are shaped by the tools and processes by which they are made [4]. Low-level tools such as PureData and Supercollider naturally afford a *bottom-up* approach described as the 'engineering default' [5]: begin with simple one-to-one linear mappings and incrementally add complexity.

A limitation of bottom-up approaches is that our representation of the body begins in the parameter space of the sensor rather than the intuitive, holistic representation we naturally have with the body, which we describe as a *semantic* representation. An example of a semantic representation includes the choreographer Laban's classification of movement by *time*, *flow*, *space*, and *weight* [6]. Much of the work of the mapping process is transforming sensor inputs into higher-level parameters, often drawing on knowledge of signal processing. Quality of movement such as 'heavy' or 'light' may be semantically simple, and one human might quickly communicate to another what they mean by these terms. But evaluating whether a sensor signal corresponds to a 'heavy' or 'light' movement is a challenge likely requiring a complex solution.

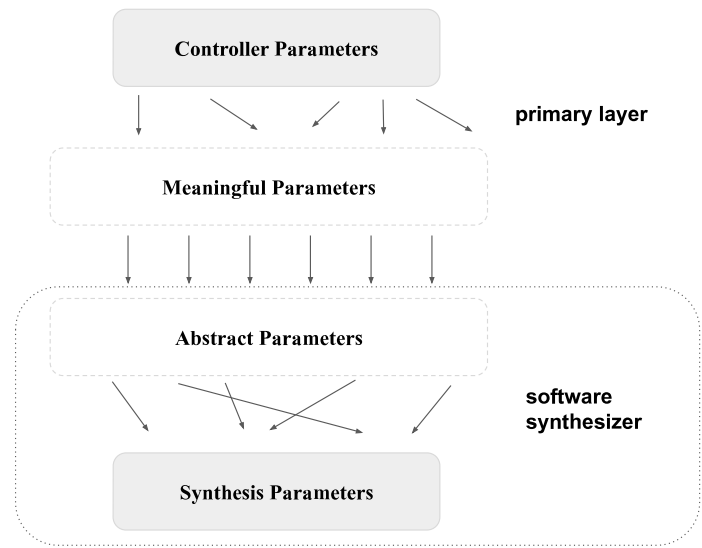

**Figure 1**
Hunt et al.'s three layer mapping model (our annotations on the right)

To overcome this distance between sensor data and semantic representation, Hunt et al. [7] propose constructing mappings in three layers, from sensor data to 'meaningful parameters' to 'abstract parameters' to synthesizer parameters (Figure 1). We consider meaningful parameters as a semantic representation of gesture, such as *weight*, and abstract parameters as a semantic representation of synthesis control, such as *brightness*.

Our work here focuses on defining this primary layer from the sensor to semantic (i.e. 'meaningful'). Complex bottom-up approaches become increasingly challenging as the complexity of sensor data increases. For example, in our case study below we use skeletal data from a Kinect v2.0, a commonly available and cheap form of motion capture. The sensor represents the body as 25 joints, each positioned in 3D space, at 30 Hz.[1] We've worked with this marvellous device (and its predecessor) for nearly a decade. In spite of extensive community contributions, our experience is that bottom-up mapping processes struggle to escape this skeletal representation: different limbs get mapped to different parameters. We are far from the intuitive meanings we might draw from Laban's analysis.

Supervised ML techniques allow creators to avoid manually defining a mapping. Instead, the creator records example gestures, labels each with a target output, and trains a model. We describe this process as *top-down*. Tools such as Wekinator [1] and MIMIC [2] support this process through Interactive Machine Learning (IML) [8] where the creator alternates between testing a model and recording training examples, allowing the creator to iteratively construct their model.

In our experience, the need to generate labelled training data, even within an IML process, imposes constraints on the creative process. It requires the creator to think in advance of which inputs they would like to associate with which outputs. However, improvising with, and composing for, an instrument, particularly one we have never encountered before, can be more a process of uncovering its affordances and constraints to reveal the instrument's character [9] [10]. Supervised ML models will naturally have their own constraints and quirks giving character to each trained model. But the process of generating training data can be challenging because the supervised paradigm is structured around solving a problem that has a ground truth. In our experience, particularly when collaborating with others, creating is less a process of marching towards our target output (what we might consider our ground truth) and more of experimenting, sensing, listening, and responding.

In observational studies on Wekinator as an IML tool, Fiebrink et al. [8] observe some composers introducing training data with the intention of generating complexity and surprise when applying the model to gestures outside of the training set. Perhaps they too do not know what mapping they are looking for at the outset. In a sense, they are 'hacking' the model by working contrary to the principles of generalisation that underlie its design. In other observational studies, West et al. [11] describe the mapping process as alternating between experimenting and exploring, and McPherson and Lepri [4] as a 'negotiation' between designer and tool. This openness suggests creators with no ground truth but instead an intuited evaluation criteria and a desire to uncover something novel.

## Open-ended mapping process

We define an *open-ended mapping process* as a process where the creator is seeking a novel, musically expressive mapping without a specific vision of what it may be, although they may hold opinions and intuitions on whether something is working or not.

We propose the following desiderata for such a process:

- *Novelty searchable*. The process should uncover novel and unexpected mappings. By novel, we mean semantically novel from the perspective of the creator in this context.
- *Ownable*. What is novel to our creator may not be novel to the scene. Creative tools have affordances and constraints that can both trap and inspire the user [4]. Musicians who create their own mapping have been observed to consider it part of their musical identity [12]. However, when encountering novelty through opaque processes such as ML, it can be unclear whether what has been created is a readily identifiable and imitable artefact of the tool (such as a synth preset) or a unique avenue worth staking our creative identity upon. Our process needs to give our creator a sense of confident authorship over the mapping.
- *Generative of expressive mappings*, which we expand on below.

Designing expressive mappings is a fundamental area of research in NIME. Key criteria have been identified including controllability, explorability, learnability [13], and diversity [14], which we might summarise as the capacity for our performer, with practice, to be able to discover and consistently reproduce a diverse range of outputs. But how much control? Which outputs? How consistently? How much practice? What is considered diverse? These are contextual and will depend on the domain, genre, composition, and performers. Nonetheless, in our multi-layered mapping model, we propose that the expressive potential of the primary layer can be maximised by the following further desiderata:

- *Consistency:* similar inputs create similar outputs.
- *Diversity:* dissimilar inputs create dissimilar outputs.
- *Range*: the entire output range can be readily generated by available inputs.

## Latent Mapping

Whereas within a supervised learning process we optimise a model to generalise from explicitly labelled data, unsupervised learning works with a corpus of unlabelled data. Instead, a model is optimised to infer some kind of structure inherent within the data itself. Such networks are designed with an architecture that requires an intermediate representation of the data, described as the *latent* representation.

We define the *latent mapping* as a function that maps gestural input to its latent representation as defined by an unsupervised model trained on a representative corpus of inputs.

The latent mapping process depends on inputs that themselves depend on the creator, project, and context at hand:

- *Input domain*: the full set of possible inputs the performer might provide
- *Model hyperparameters*, such as the number of output dimensions.

## Generating Latent Mappings using Variation Autoencoders (VAEs)

We propose that the Variational Autoencoder (VAE) [15] when used as a latent mapping satisfies our desiderata.

The VAE is an unsupervised learning method that uses deep neural networks to encode and decode data to a lower-dimensional space of latent representations. It is effectively a lossy form of compression created bespoke to the dataset which maximizes the model's ability to reconstruct the original input. In our example of skeletal pose data, a single frame of 75 numbers is *encoded* into a latent space of 16 gaussian distributions and then *decoded* back into a reconstructed skeleton (Figure 2). The *encoder* and *decoder* capture information about what is consistent across the dataset, while the latent variable captures what is different about this specific input, given those identified consistencies.

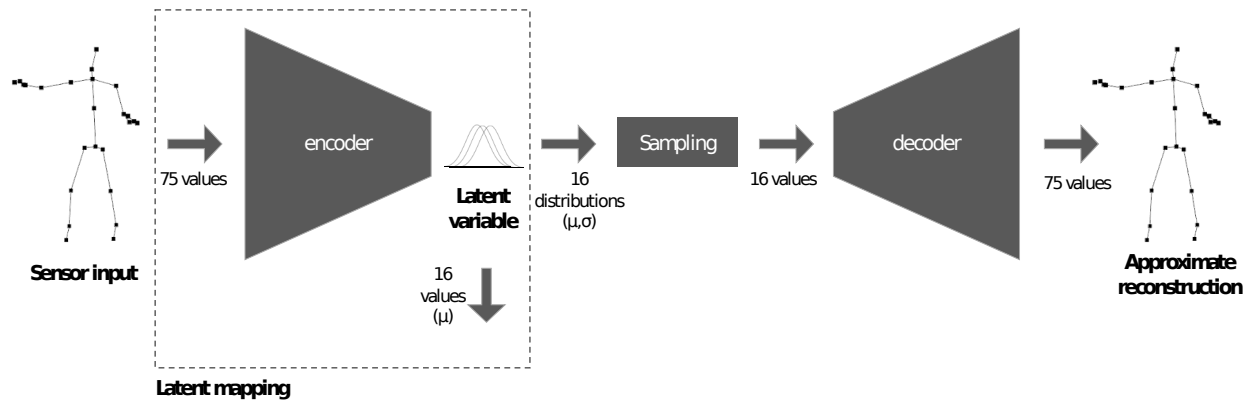

**Figure 2**
A variational autoencoder, highlighting the section used for the latent mapping.

In our latent mapping implementation, we take as output the 16 parameters defining the means of these distributions.

The VAE is often used for generating examples similar to the training set by taking the decoder to decode samples from the latent space. For example, Crnkovic-Friis and Crnkovic-Friis [16] and Pettee et al. [17] use an Autoencoding Recurrent Neural Network (RNN) to generate novel choreography from a corpus of motion-capture data from a dancer. Pettee et al. [18] train a Graph Neural Network to generate choreographies in real-time responding to a dancer. By contrast, our approach uses the encoder side as a primary mapping function from sensor input to a lower-dimensional latent space. In many cases, dimensions in the latent space have been observed to correspond to semantically meaningful qualities, such as whether a face is smiling or a hand-written digit is rotated [19].

During training, the VAE's decoder samples from the latent space represented as gaussian distributions. This introduces noise requiring the decoder to map similar latent representations to similar reconstructions, which enforces smoothness on the decoder and encoder (*Consistency*).[2] This smoothness combined with the overall penalty on reconstruction error implies non-similar inputs will have non-similar latent representations (*Diversity*). To satisfy our *Range* desideratum, we additionally apply a per-component normalization on the latent representations (post-training) to ensure full coverage of the space [0,1][16]. To normalize, we linearly map either the entire range of ±2 standard deviations to the range [0,1], clipping if necessary.

The random initialisation of weights means a novel mapping is generated each time it is trained (*Novelty searchable*). Nonetheless, the creator retains a sense of authorship by training exclusively on their own gestural vocabulary (*Ownable*).

## Case Study: Sonified Body

We present as a case study our own artistic explorations using this approach in a project entitled *Sonified Body*.

Our dataset was 16 hours of skeletal movement, recorded by the first author over six weeks as a daily practice of 30 minutes of improvised movement. The recording occurred at the artist's studio space using a Kinect v2.0, giving 75-dimensional input at 30 Hz. We trained a VAE with a 16-dimensional latent space with this data to generate a latent mapping.

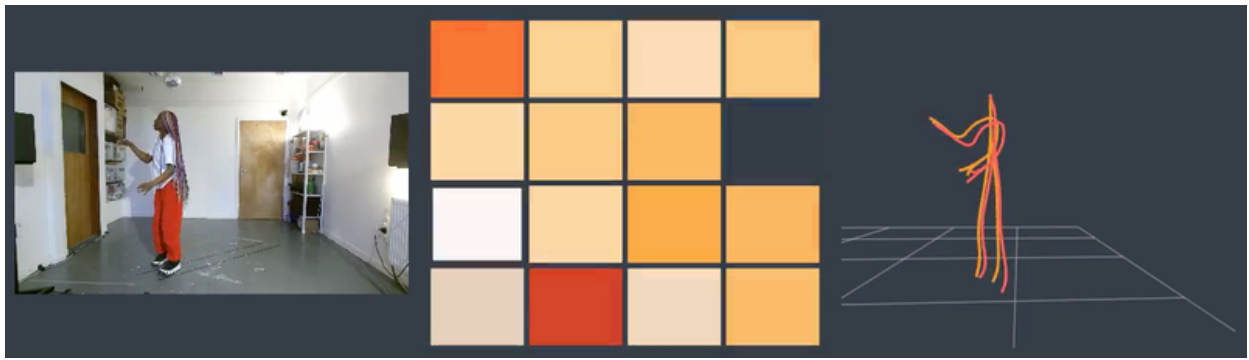

**Figure 3**
A screengrab from our real-time performance system showing a reference RGB camera feed, the input (orange) and reconstructed skeleton (red) and a heatmap illustrating the current latent variable.

We integrated this model into a real-time performance system (Figure 3), which outputted OSC to Max for Live patches that modulated visible parameters of the built-in presets of the bundled software synthesizers in Ableton Live (Figure 4). As our focus was on understanding the qualities of the VAE's mapping, we kept this second mapping layer somewhat arbitrary. However, we avoided parameters such as a master gain control that might negate the effect of many other parameters.

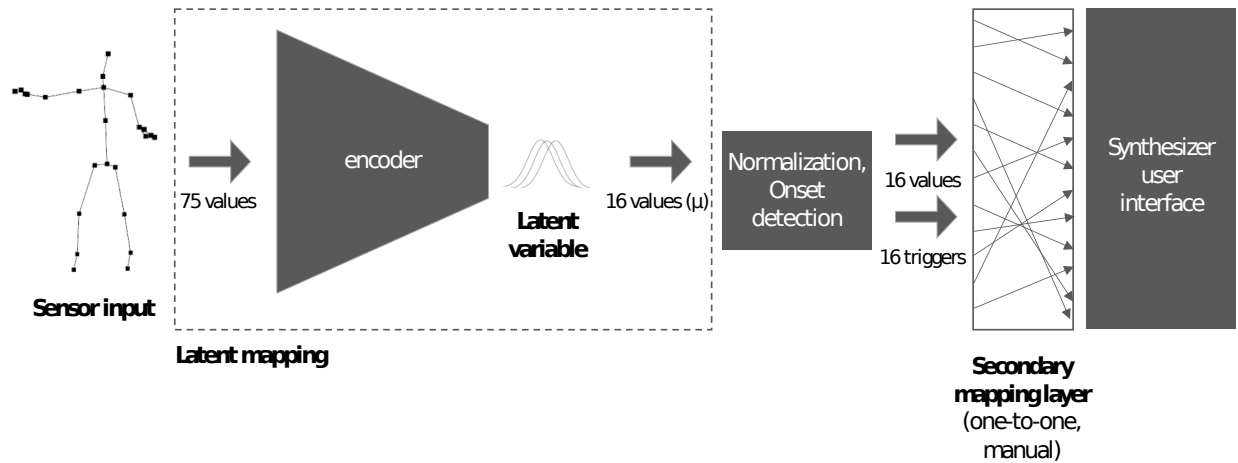

**Figure 4**
The *Sonified Body* system.

The mapping layer currently takes around 2 milliseconds on a GPU and remains under 20 ms on a CPU. Combined with a reported 67 ms latency from the Kinect [20] and 5 ms audio processing buffer gives a total latency of around 75 ms, which is suitable for our current artistic needs. A different sensor could significantly reduce this value.

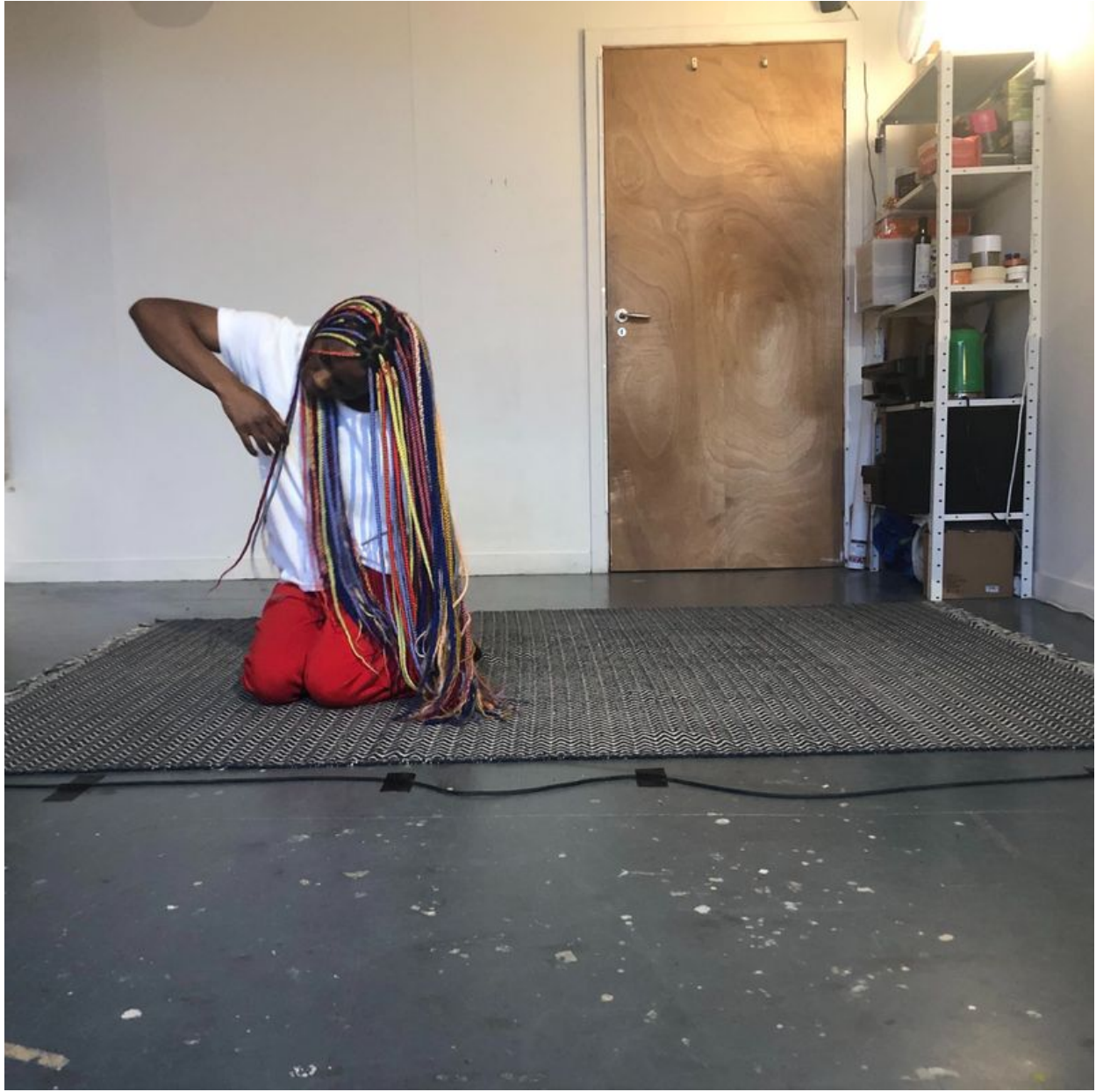

**Figure 5**
The dancer Divine Tasinda improvising with the system during the exploratory residency.

## Exploration with dancers

During a 5-day artist residency (spread over 2 months due to pandemic restrictions) we explored the system with three dancers, each from a different movement background but experienced in improvisation (Figure 5).

After the first two days of the lab, we felt the need for discrete events and developed an onset detection algorithm based loosely on Dahl et al.'s [21] technique of identifying crossing points of two low pass filters. We applied this to each output component on the latent space giving 16 potential trigger events, which we mapped to drum kit presets.

We felt the results were strong enough to record a number of improvised performances that were streamed as part of the Present Futures online festival.

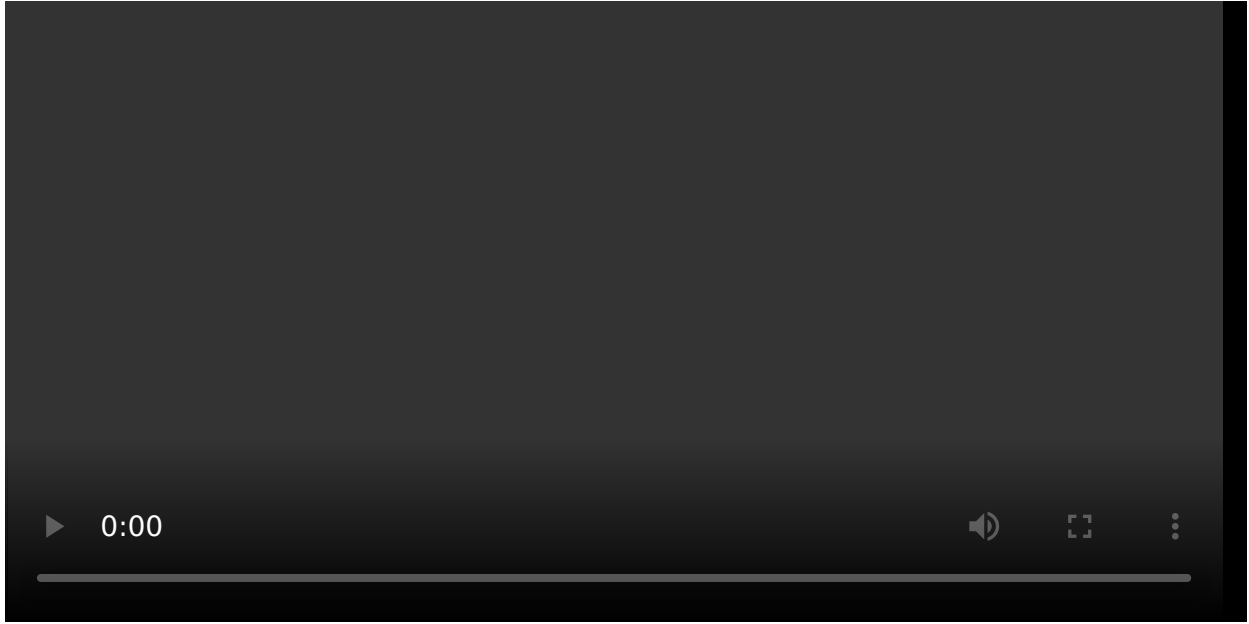

**Video 1**

Improvised performances with the sonified body system by dancers Catriona Robertson, Divine Tasinda and Adilso Machado. Video also available at [https://timmb.com/sonified-body-r-and-d-lab](https://timmb.com/sonified-body-r-and-d-lab).

Our preliminary observations were that each dancer found a distinctive way of performing with the system, drawing on their own diverse movement languages. At the same time, they did each adapt their movement as they became more familiar with the system. For example, we noticed a tendency towards 'Kinect-friendly' gestures, such as focusing on limb extension above facial expression.

We had between 5 and 14 hours with each dancer. All permutations of synthesizers and model settings were prepared in advance. No training was done during the residency itself, giving us ample time to test these permutations and record a number of performances.

## Discussion

We have argued that for some creators, the creation of mapping is an open-ended creative process. This may not readily fit with current tools which support either manual definition of mapping functions or their derivation through supervised learning. We present this position as a desideratum, argued from both existing research and our own practices, that suggests latent mappings may be more appropriate in these cases. Just as there is no universal creative process, there can be no universal approach to creating mappings. We have sought to qualify our approach in terms of the creative needs that a creator may or may not identify with.

The encoder half of a VAE is a latent mapping with desirable properties. Other model architectures may provide interesting alternatives. For example, an RNN's [22][23] latent representation retains a memory of previous input, which could capture how a gesture is changing through time.

So far we have focused on the primary layer of a greater mapping. In our case study, we mapped the latent mapping arbitrarily to the user-friendly parameters presented by a number of commercial software synthesisers. Encouraged by these results, we are interested in applying similar unsupervised approaches to sound synthesis.

Our VAE-based latent mapping outputs continuous parameter values. To trigger note events, we deployed bottom-up approaches in the subsequent mapping layer. This introduced context-specific 'magic numbers' that we had hoped machine learning would relieve us from finding. We plan to explore techniques to extract discrete events directly from the model.

Dataset diversity and prejudicial bias is a critical issue to explore when training models on humans [24]. Our findings suggest the potential for one individual to generate enough skeletal joint data to train a latent mapping that can generalise to new inputs from that individual. In the case where a single individual both trains and performs with the system then there is no need for generalisation to other individuals and so dataset bias is not an issue. In fact, generalisation may be undesirable as it may reduce that individual's sense of ownership.

However, if other performers are to use the system then this is no longer the case and ethical consideration is needed to ensure the model does not capture a prejudicial bias. In our case study, the model is trained on a White male and then tested with a White male, a White female, and a Black female with the potential consequences of this as a line of critical artistic enquiry. At this stage, we can report anecdotally that our system does not appear to show differences in behaviour between these individuals. However, as we are working with the Kinect's skeleton representation then we are piggy-backing on whatever work Microsoft has done to minimise bias in the sensor's capabilities. If our input were raw camera data, we would not expect similar generalisability. In future work, we plan to investigate how our desiderata may be quantified statistically giving one tool towards investigating this potential bias more rigorously.

## Acknowledgements

*Sonified Body* features the dancers Adilso Machado, Divine Tasinda, and Catriona Robertson. It was mentored by Ghislaine Boddington and produced by Feral. It was funded by Creative Scotland and Present Futures festival with support from Preverbal Studio.

## Compliance with Ethical Standards

The dancers invited to the residency provided informed consent for their involvement and the published outputs. They were paid for their time at rates above the minimum set by the Independent Theatre Council.

## Footnotes

1.  Arguably, this representation is already meaningful given the sensor's initial representation as a time-of-flight infrared image. Microsoft provides drivers that transparently infer the 25 joint skeleton. For the purposes of our argument, we consider this skeletal representation as our raw sensor input. ↩

2.  This smoothness further depends on a regularization parameter that penalises the Kullback-Leibler divergence from the uniform distribution. This ensures the model does not remove the noise by minimising the variance in the Gaussians of its latent representation. ↩

## Citations

1. Fiebrink, R., Trueman, D., & Cook, P. R. (2009). A Meta-Instrument for Interactive, On-the-fly Machine Learning. In *Proceedings of the conference on New Interfaces for Musical Expression*. Pittsburg, PA. ↩
2. McCallum, L., & Grierson, M. S. (2020). Supporting Interactive Machine Learning Approaches to Building Musical Instruments in the Browser. In R. Michon & F. Schroeder (Eds.), *Proceedings of the International Conference on New Interfaces for Musical Expression* (pp. 271–272). Birmingham, UK: Birmingham City University. ↩